\documentclass[longauth]{aa} 
\usepackage{graphicx}
\usepackage{txfonts}
\usepackage{epsfig}
\usepackage{longtable}
\usepackage{lscape}
\usepackage{amssymb}
\usepackage[colorlinks=true,citecolor=blue]{hyperref} 
\usepackage{refcount}

\bibpunct{(}{)}{;}{a}{}{,} 

\def\src {4C~$+$71.07}

\def \sw {{\em Swift}}

\def \ergsec{\hbox{erg s$^{-1}$}}
\def \ferg {erg cm$^{-2}$ s$^{-1}$}
\def \hcm {\hbox {\ifmmode $ atom cm$^{-2}\else atom cm$^{-2}$\fi}}

\def \arcsec {\hbox{$^{\prime\prime}$}}

\def \agile {AGILE}

\def \fermi{{\it Fermi}}

\def \phcmsec{\hbox{photons cm$^{-2}$ s$^{-1}$}}
\def \gray {$\gamma$-ray}

\def \apj {ApJ}
\def \apjl {ApJL}
\def \apjs {ApJS}
\def \aap {A\&A}

\def \mnras {MNRAS}

\def \ssr {SSRv}

\begin{document}

   \title{AGILE, \emph{Fermi}, \emph{Swift}, and GASP/WEBT\thanks{Partly based on data taken and assembled by 
   the WEBT collaboration and stored in the WEBT archive at the Osservatorio Astrofisico di 
   Torino -- INAF (\href{http://www.oato.inaf.it/blazars/webt}{http://www.oato.inaf.it/blazars/webt}).}  
   multi-wavelength observations of  the high-redshift blazar 4C~$+$71.07 in outburst}
\titlerunning{Multi-wavelength observations of \src{} in outburst}
\authorrunning{Vercellone et al.}

  \author{S.~Vercellone\inst{\ref{oab}}
  \and P.~Romano\inst{\ref{oab}}
  \and G. ~Piano\inst{\ref{iaps}}
  \and V.~Vittorini\inst{\ref{iaps}}
  \and I.~Donnarumma\inst{\ref{asi}}
  \and P. ~Munar-Adrover\inst{\ref{uab}, \ref{iaps}}
  \and C.~M.~Raiteri\inst{\ref{oato}}
  \and M.~Villata\inst{\ref{oato}}
  \and F.~Verrecchia\inst{\ref{ssdc},\ref{oarm}}
  \and F.~Lucarelli\inst{\ref{ssdc},\ref{oarm}}
  \and C.~Pittori\inst{\ref{ssdc},\ref{oarm}}
  \and A.~Bulgarelli\inst{\ref{iasfbo}}
  \and V.~Fioretti\inst{\ref{iasfbo}}
  \and M.~Tavani\inst{\ref{iaps},\ref{tov},\ref{gssi}}
  \and J.~A.~Acosta-Pulido\inst{\ref{IAC}, \ref{UniLaguna}}
  \and I.~ Agudo\inst{\ref{CSIC}}
  \and A.~A.~Arkharov\inst{\ref{Pulkovo}}
  \and U.~Bach\inst{\ref{mpifr}}
  \and R.~Bachev\inst{\ref{Bulgarian}}
  \and G.~A.~Borman\inst{\ref{Crimean}}
  \and M.~S.~Butuzova\inst{\ref{Crimean}}
  \and M.~I.~Carnerero\inst{\ref{oato}}
  \and C.~Casadio\inst{\ref{mpifr}}
  \and G.~Damljanovic\inst{\ref{AOVolgina}}
  \and F.~D'Ammando\inst{\ref{ira}, \ref{unibo}}
  \and A.~Di Paola\inst{\ref{oarm}}
  \and V.~T.~Doroshenko\thanks{Deceased.}\inst{\ref{Sternberg}}
  \and N.~V.~Efimova\inst{\ref{Pulkovo}}
  \and Sh.~A.~Ehgamberdiev\inst{\ref{Ulugh}}
  \and M.~Giroletti\inst{\ref{ira}}
  \and J.~L.~G\'omez\inst{\ref{CSIC}}
  \and T.~S.~Grishina\inst{\ref{StPetersburg}}
   \and E.~J\"arvel\"a\inst{\ref{Metsahovi}, \ref{Aalto}}
  \and S.~A.~Klimanov\inst{\ref{Pulkovo}}
  \and E.~N.~Kopatskaya\inst{\ref{StPetersburg}}
  \and O.~M.~Kurtanidze\inst{\ref{Abastumani}, \ref{Engelhardt}, \ref{Guangzhou}, \ref{KeyLab}}
  \and A. L\"ahteenm\"aki\inst{\ref{Metsahovi}, \ref{Aalto}}
  \and V.~M.~Larionov\inst{\ref{Pulkovo}, \ref{StPetersburg}}
  \and L.~V.~Larionova\inst{\ref{StPetersburg}}
  \and B.~Mihov\inst{\ref{Bulgarian}}
  \and D.~O.~Mirzaqulov\inst{\ref{Ulugh}}
  \and S.~N.~Molina\inst{\ref{CSIC}}
  \and D.~A.~Morozova\inst{\ref{StPetersburg}}
  \and S.~V.~Nazarov\inst{\ref{Crimean}}
  \and M.~Orienti\inst{\ref{ira}}
  \and S.~Righini\inst{\ref{ira}}
  \and S.~S.~Savchenko\inst{\ref{StPetersburg}}
  \and E.~Semkov\inst{\ref{Bulgarian}}
  \and L.~Slavcheva-Mihova\inst{\ref{Bulgarian}}
  \and A.~Strigachev\inst{\ref{Bulgarian}}
  \and M.~Tornikoski\inst{\ref{Metsahovi}}
  \and Yu.~V.~Troitskaya\inst{\ref{StPetersburg}}
  \and O.~Vince\inst{\ref{AOVolgina}}
  \and P.~W.~Cattaneo\inst{\ref{infnpv}}
  \and S.~Colafrancesco\thanks{Deceased.}\inst{\ref{wits}}
  \and F.~Longo\inst{\ref{infnts},\ref{units}}
  \and A.~Morselli\inst{\ref{infnToV}}
  \and F.~Paoletti\inst{\ref{RSD},\ref{iaps}}
  \and N.~Parmiggiani\inst{\ref{iasfbo}}
}
\institute{INAF, Osservatorio Astronomico di Brera, Via Emilio Bianchi 46, I-23807 Merate (LC), Italy \\
  \email{stefano.vercellone@inaf.it}\label{oab}
\and INAF, Istituto di Astrofisica e Planetologia Spaziale, Via Fosso del Cavaliere 100, I-00133 Roma, Italy \label{iaps}
\and ASI, Via del Politecnico, I-00133 Roma, Italy \label{asi}
\and UAB, Universitat Auton\`{o}ma de Barcelona, Departament de F\'{i}sica Edifici C, 08193 Bellaterra (Cerdanyola del Vall\`{e}s), Spain \label{uab}
\and INAF, Osservatorio Astronomico di Torino, Via Osservatorio 20, I-10025 Pino Torinese, Italy \label{oato}
\and ASI Space Science Data Center, Via del Politecnico, I-00133 Roma, Italy \label{ssdc}
\and INAF, Osservatorio Astronomico di Roma, via Frascati 33, I-00078 Monte Porzio Catone, Italy \label{oarm}
\and INAF, Osservatorio di Astrofisica e Scienza dello Spazio, Via Piero Gobetti 93/3, I-40129 Bologna, Italy \label{iasfbo}
\and Dip. di Fisica, Univ. di Roma ``Tor Vergata'', Via della Ricerca Scientifica 1, I-00133 Roma, Italy \label{tov}
\and Gran Sasso Science Institute, viale Francesco Crispi 7, I-67100 L'Aquila, Italy \label{gssi}
\and Instituto de Astrof\'{\i}sica de Canarias (IAC), E-38205 La Laguna, Tenerife, Spain \label{IAC}
\and Departamento de Astrof\'{\i}sica, Universidad de La Laguna, E-38206 La Laguna, Tenerife, Spain \label{UniLaguna}
\and Instituto de Astrof\'{\i}sica de Andaluc\'{\i}a (CSIC), Apartado 3004, E-18080 Granada, Spain \label{CSIC}
\and Pulkovo Observatory, St.-Petersburg, Russia \label{Pulkovo}
\and Max-Planck-Insitut f\"ur Radioastronomie, Auf dem H\"ugel 69, 53121 Bonn, Germany \label{mpifr}
\and Institute of Astronomy and National Astronomical Observatory, Bulgarian Academy of Sciences, 72, Tsarigradsko Shose Blvd., 1784 Sofia, Bulgaria \label{Bulgarian}
\and Crimean Astrophysical Observatory RAS, p/o Nauchny, 298409, Russia \label{Crimean}
\and Astronomical Observatory, Volgina 7, 11060 Belgrade, Serbia \label{AOVolgina}
\and INAF, Istituto di Radioastronomia, via Piero Gobetti 93/2, I-40129 Bologna, Italy \label{ira}
\and Dip. di Fisica e Astronomia, Universit\`a di Bologna, Viale Berti Pichat 6/2, I-40127 Bologna, Italy \label{unibo}
\and Sternberg Astronomical Institute, M.V.Lomonosov Moscow State University, Universitetskij prosp.13, Moscow 119991, Russia \label{Sternberg}
\and Ulugh Beg Astronomical Institute, Maidanak Observatory, Uzbekistan \label{Ulugh}
\and Astron.\ Inst., St.-Petersburg State Univ., Russia \label{StPetersburg}
\and Aalto University Mets\"ahovi Radio Observatory, Mets\"ahovintie 114, 02540 Kylm\"al\"a, Finland \label{Metsahovi}
\and Aalto University Department of Electronics and Nanoengineering, P.O. BOX 13000, FI-00076 Aalto, Finland \label{Aalto}
\and Abastumani Observatory, Mt. Kanobili, 0301 Abastumani, Georgia \label{Abastumani}
\and Engelhardt Astronomical Observatory, Kazan Federal University, Tatarstan, Russia \label{Engelhardt}
\and Center for Astrophysics, Guangzhou University, Guangzhou 510006, China \label{Guangzhou}
\and Key Laboratory of Optical Astronomy, National Astronomical Observatories, Chinese Academy of Sciences, Beijing 100012, China \label{KeyLab}
\and INFN, sezione di Pavia, Via Agostino Bassi, 6, I-27100 Pavia, Italy \label{infnpv}
\and School of Physics, University of the Witwatersrand, 1 Jan Smuts Avenue, Braamfontein 2000 Johannesburg, South Africa \label{wits}
\and INFN, sezione di Trieste, via Valerio 2, I-34127 Trieste, Italy \label{infnts}
\and Dipartimento di Fisica, Universit\`{a} degli Studi di Trieste, via Valerio 2, I-34127 Trieste, Italy \label{units}
\and INFN, sezione di Roma Tor Vergata, via della Ricerca Scientifica 1, I-00133 Roma, Italy \label{infnToV}
\and East Windsor RSD, 25a Leshin Lane, Hightstown, NJ 08520, USA \label{RSD}
}

\date{Received...; accepted... }
 
\abstract{
The flat-spectrum radio quasar \src{} is a high-redshift ($z=2.172$), $\gamma$-loud blazar whose optical emission is dominated
by the thermal radiation from accretion disc.
}
{
\src{} has been detected in outburst twice by the AGILE \gray{} satellite during the period end of October -- mid November 2015,  when it reached
a \gray{} flux of the order of $F_{\rm E>100\,MeV} = (1.2 \pm 0.3)\times 10^{-6}$\,\phcmsec and 
$F_{\rm E>100\,MeV} = (3.1 \pm 0.6)\times 10^{-6}$\,\phcmsec,
respectively,  allowing us to investigate the properties of the jet and of the emission region.
}
{
We investigated its spectral energy distribution by means of almost simultaneous observations covering the cm, mm, near-infrared, optical,
ultra-violet, X-ray and \gray{} energy bands obtained by the GASP-WEBT Consortium, the {\it Swift} and the AGILE and {\it Fermi} satellites.
}
{
The spectral energy distribution of the second \gray{} flare (the one whose energy coverage is more dense) can be modelled by means 
of a one-zone leptonic model,  yielding a total jet power of about $4\times10^{47}$\,erg\,s$^{-1}$.
}
{
During the most prominent \gray{} flaring period our model is consistent with a dissipation region within the broad-line region.  Moreover, this
class of high-redshift, large-mass black-hole flat-spectrum radio quasars might be good targets for future \gray{} satellites such as e-ASTROGAM.
}

\keywords{Flat-spectrum radio quasar objects: individual: \src\ -- galaxies: active -- X-rays: individual: \src.  }

   \maketitle
%

  \section{Introduction \label{s0836:intro}}

Among the active galactic nuclei (AGNs), blazars show the most extreme properties. 
Their variable emission spans several decades of energy, from the radio to the TeV
energy band, with variability time-scales ranging from a few minutes, such as 
PKS~2155$-$304~\citep[][]{2007ApJ...664L..71A} and 3C~279~\citep[][]{2016ApJ...824L..20A}, 
up to a few years \citep[e.g., BL Lacertae][]{2013MNRAS.436.1530R}.
Radio observations often reveal superluminal motion and brightness temperatures exceeding the Compton limit. 
These features can be explained by assuming that blazars emit mainly non-thermal radiation (synchrotron at low 
energies and inverse-Compton at high energies) coming from a relativistic plasma jet oriented close to the line of sight, 
with consequent Doppler beaming \citep[e.g., ][]{1995PASP..107..803U}. 
Therefore, the study of blazar emission is mainly an investigation of the properties of plasma jets in AGNs. 
The blazar class includes flat-spectrum radio quasars (FSRQs) and BL Lac objects. FSRQs show evidence 
of additional unbeamed emission contributions from the nucleus, i.e. thermal radiation from the accretion disc 
in the rest-frame UV, and broad emission lines due to fast-rotating gas clouds surrounding the disc 
(the so-called broad-line region, BLR).

The flat-spectrum radio quasar \src{} \citep[S5~0836$+710$;  $z = 2.172$,][]{1993A&AS..100..395S} is known as a \gray{} emitter since
its detection by the Energetic Gamma Ray Experiment Telescope (EGRET) instrument on board the
Compton Gamma Ray Observatory (CGRO) \citep{1999ApJS..123...79H}, and it is one of the $\gamma$-loud blazars
monitored by the GLAST-AGILE Support Program (GASP) of the Whole Earth Blazar Telescope (WEBT) 
Collaboration \citep{2008A&A...481L..79V}.\footnote{\href{http://www.oato.inaf.it/blazars/webt}{http://www.oato.inaf.it/blazars/webt}.} 
This allows us to study its multi-wavelength flux behaviour on a long time scale.
In the last seven years the optical and millimetre light curves have shown several flares that however do not seem 
to be correlated, at least not in a straightforward way.
Moreover, the correlation between the optical flux variations and those observed in the \gray{} energy band 
by the {\em Fermi} satellite appears to be complex.
Complex correlations between emission in various bands may reflect unusual processes and/or a jet structure in this source that 
make a detailed multifrequency study worthwhile~\citep[see][for a long-term, multi-wavelength monitoring study 
of this source]{2013A&A...556A..71A}. 
Another peculiarity of \src{} is its relatively high redshift (it is the most distant FSRQ in the GASP--WEBT sample) 
and an intervening system at $z=0.914$ \citep{1993A&AS..100..395S}. 
Intervening systems producing Mg~II $\lambda \lambda$2796,2803$\AA$ absorption lines in the AGN spectrum may be due 
to a chance alignment of, e.g., a galaxy with respect to the AGN light of sight.

The spectral energy distribution (SED) of \src{} presents a double-hump morphology typical of blazars, with a 
synchrotron peak in the far-infrared band and an inverse Compton peak at about $10^{20}$--$10^{21}$\,Hz 
\citep{2007ApJ...669..884S, 2013A&A...556A..71A}.
Moreover, it shows a strong blue bump peaking at about $10^{14.9}$\,Hz, which is the signature of an accretion disc,
whose luminosity is comparable to the highest values observed in type 1 QSO \citep{2014MNRAS.442..629R}.

\src{} was detected in a flaring state by AGILE at the end of October and at the beginning of November 2015, 
\citep[][]{2015ATel.8223....1B, 2015ATel.8266....1P} and followed-up by the Neil Gehrels {\it Swift} Observatory
\citep[][]{2015ATel.8229....1V, 2015ATel.8271....1V} and the GASP--WEBT. 
These observations allow us to investigate its SED during both \gray{} flares by means of almost simultaneous data, and
to study their possible modulations and correlations.

This paper is organised as follows. In Sections~\ref{s0836:data} we present the AGILE, {\it Fermi}-LAT, \sw{}, XMM-Newton and GASP-WEBT 
data analysis and results. In Section~\ref{s0836:discussion}, we present the simultaneous multi-wavelength light curves, 
the SEDs of the two \gray{} flares, and we discuss the results. In Section~\ref{s0836:conclusions} we draw our conclusions.
Throughout this paper the quoted uncertainties are given at the 1$\sigma$ level, unless otherwise stated, and we adopt
a $\Lambda$CDM cosmology~\citep{2016A&A...594A..13P}.

   \section{Observations and data analysis \label{s0836:data}}

                \subsection{AGILE \label{s0836:data_agile}}
%
%
The AGILE satellite \citep{Tavani2009:Missione} is a mission of the Italian Space Agency (ASI) devoted to 
high-energy astrophysics. 
The AGILE scientific instrument combines four active detectors yielding broad-band coverage from hard X-ray
to \gray{} energies: a Silicon Tracker~\citep[ST;][30~MeV--50~GeV]{Prest2003:agile_st},
a co-aligned coded-mask hard X-ray imager, Super--AGILE \citep[SA;][18--60~keV]{Feroci2007:agile_sa}, 
a non-imaging CsI Mini--Calorimeter~\citep[MCAL;][0.3--100~MeV]{Labanti2009:agile_mcal},
and a segmented Anti-Coincidence System~\citep[ACS;][]{Perotti2006:agile_ac}.
\gray{} detection is obtained by the combination of ST, MCAL and ACS; these three detectors form the 
AGILE Gamma-Ray Imaging Detector (GRID).
A ground segment alert system allows the AGILE team to perform the full AGILE-GRID data reduction 
and the preliminary quick-look scientific analysis \citep{2013NuPhS.239..104P,2014ApJ...781...19B}.

   \begin{figure}
   \resizebox{\hsize}{!}{\includegraphics[angle=-90]{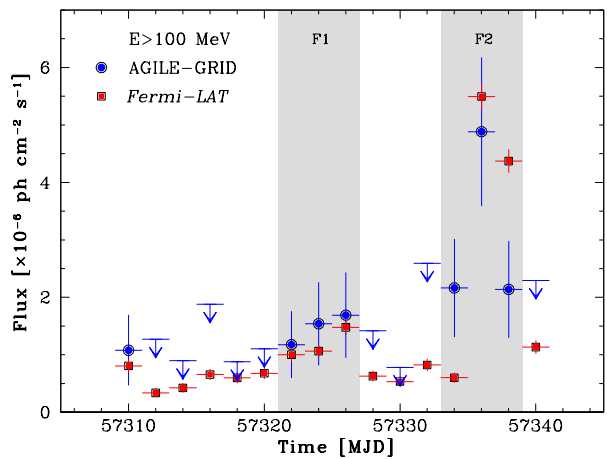}}
   \caption{AGILE-GRID (blue circles) and {\it Fermi}-LAT (red squares) light-curves (48h-bins) covering the period 
   2015 October 15 -- November 15. Downward arrows represents 
   2$\sigma$ upper-limits. The grey-dashed areas mark the time-interval used for accumulating the almost simultaneous SEDs.}
    \label{s0836:fig:agilelc}
    \end{figure}

AGILE-GRID data were analysed by means of the AGILE standard analysis pipeline 
\citep[see][for a description of the AGILE data reduction]{Vercellone2008:3C454_ApJ}
and the standard {\tt FM3.119} AGILE filter.
The \gray{} counts, exposure and diffuse emission maps, needed for the
analysis were created with a bin size of $0.5^\circ \times 0.5^\circ$, 
energy E $>$ 100 MeV, and off-axis angles lower than $50^\circ$.
We rejected all the \gray{} events within a cone of $90^\circ$ half-opening angle 
with respect to the satellite--Earth vector, in order to reduce the \gray{}
Earth albedo contamination. 
We used the latest version (BUILD-23) of the
calibration matrices (I0025), with the updated version of the \gray{}
diffuse emission model \citep{Giuliani2004:diff_model}. 
A multi-source maximum likelihood analysis \citep[ALIKE, ][]{2012A&A...540A..79B}
based on the Test Statistic method as formulated by \cite{1996ApJ...461..396M}
was carried out with an analysis radius of $10^\circ$ and GAL-ISO parameters
(indicating the relative weights of the Galactic and isotropic diffuse components)
fixed at the values calculated during the two weeks preceding the analyzed AGILE dataset (2015 October 1--14).

In order to produce the AGILE light-curve (see Fig.~\ref{s0836:fig:agilelc}), we divided the data collected in the period
2015 October 14 -- November 15 (MJD: 57309.0 -- 57341.0) in 48h-bins.
The ALIKE was carried out by fixing the position of the source to its nominal one (l, b) = ($143.54^\circ$, $34.43^\circ$), 
 \cite{Myers2003}.
We have performed the spectral analysis of the activity at the two \gray{} peaks, corresponding to the
periods between  2015-10-26 (MJD: 57321.0) and  2015-11-01 (MJD: 57327.0),
and between 2015-11-07 (MJD: 57333.0) and 2015-11-13 (MJD: 57339.0), 
labelled flare--1 (F1) and flare--2 (F2), respectively, and marked by means of grey-dashed areas.
Moreover, we accumulated the average spectrum integrating between 2015-10-14 (MJD: 57309.0) 
and 2015-11-15 (MJD: 57341.0).

 \begin{table} 	
 \begin{center} 	
  \scriptsize
\caption{AGILE-GRID and {\it Fermi}-LAT \gray{} fluxes and spectral indices. All dates are at 00:00:00 UTC.} 	
 \label{s0836:tab:fgamma} 	
 \begin{tabular}{ lllll } 
 \hline 
 \hline 
 \noalign{\smallskip} 
  Label    & T$_{\rm start}$  & T$_{\rm stop}$ & F$_{\rm E>100\, MeV}$$^{a}$      & $\Gamma_{\gamma}$   \\ 
 \noalign{\smallskip} 
   \hline 
 \noalign{\smallskip} 
%
%
\multicolumn{5}{c}{{\it AGILE-GRID}}\\
   F1                        & 2015-10-26              & 2015-11-01            &$(1.2 \pm 0.3)$		          &  $(1.7 \pm 0.5)$  \\
   F2                        & 2015-11-07              & 2015-11-13            &$(3.1 \pm 0.6)$		          &  $(2.3 \pm 0.3)$  \\
   Whole period       & 2015-10-14              & 2015-11-15            &$(0.94\pm 0.15)$		         &  $(2.1 \pm 0.2)$  \\
\hline
%
%
\multicolumn{5}{c}{{\it Fermi-LAT}}\\
   F1                        & 2015-10-26              & 2015-11-01            &$(1.18 \pm 0.07)$		          &  $(2.72 \pm 0.08)$  \\
   F2                        & 2015-11-07              & 2015-11-13            &$(3.4 \pm 0.1)$		          &  $(2.54 \pm 0.04)$  \\
   Whole period       & 2015-10-14              & 2015-11-15            &$(1.30 \pm 0.03)$		        &  $(2.64 \pm 0.03)$  \\
\noalign{\smallskip}
  \hline
  \end{tabular}
  \end{center}
  {\footnotesize $^{a}$: Fluxes ($E>100$\,MeV) in units of $10^{-6}$\,\phcmsec{}.}
  \end{table} 

Table~\ref{s0836:tab:fgamma} shows the \gray{} fluxes obtained by integrating in the whole AGILE energy band
(100\,MeV -- 50\,GeV) and the photon indices obtained by a fit with a power-law.
We restricted the \gray{} spectral analysis to three energy bins: 100--200, 200--400, and 400--1000\,MeV.  
The \gray{} photon indices are consistent within the errors.

                \subsection{Fermi-LAT \label{s0836:Fermi}}
                
\src{} data were retrieved using the {\it Fermi} data access 
service\footnote{{\tt http://fermi.gsfc.nasa.gov}}. We selected PASS8 data centered at the position of the source 
with a radius of $25^{\circ}$ covering the same time-interval as the AGILE data. We analyzed the data using the 
{\it Fermi} Science Tools version {\tt v11r5p3} and with the P8R2\_SOURCE\_V6 instrument response function 
(IRF)\footnote{For more information about IRFs and details, the reader is referred to the 
{\it Fermi} instrumental publications}. In order to analyze the data we also made use of the user contributed package 
{\tt Enrico}\footnote{{\tt https://github.com/gammapy/enrico/}}.

In our work, we adopted the current Galactic diffuse emission model ({\tt gll\_iem\_v06.fits}) and also the current 
isotropic emission model ({\tt iso\_P8R2\_SOURCE\_V6\_v06.txt}) within the likelihood analysis. 
We took into account nearby sources that are present in the {\it Fermi}-LAT 3rd point source catalog  
\citep[{\tt gll\_psc\_v16.fit},][]{2015ApJS..218...23A}. We selected an energy range between 100~MeV and 300~GeV 
and filtered the events for the source class. We limited the reconstructed zenith angle to be less than 
90$^{\circ}$ to greatly reduce gamma rays coming from the limb of the Earth's atmosphere. 
We selected the good time intervals of the observations by excluding events that were taken 
while the instrument rocking angle was larger than 52$^{\circ}$.

Our analysis was done in two steps: in the first step, all sources within $10^{\circ}$ angular distance to our source 
of interest had their spectral parameters free, while the sources at greater distance and up to $25^{\circ}$, 
had their parameters fixed. A likelihood analysis was performed to fit the parameters using a {\tt Minuit} optimizer. 
In a second step, we fixed the nearby sources parameters to the ones fitted in the previous step and run again 
the likelihood analysis, with the {\tt NewMinuit} optimizer. In both steps, our source of interest had its parameters 
free and both Galactic diffuse and isotropic emission parameters were fixed. 

These steps were performed as explained in three different analysis: one for the whole period, 
one for the first gamma-ray flare, and one for the second gamma-ray flare.

                \subsection{XMM-Newton \label{s0836:data_xmm}}
%
%
In order to best constrain the uncertainty in the absorption as derived from the relatively short \sw{} observations, 
we reanalysed archival {\em XMM-Newton} observations with the Science Analysis 
Software\footnote{\href{http://xmm.esac.esa.int/sas/}{http://xmm.esac.esa.int/sas/}} (SAS) version 14.0.0 following 
standard prescriptions and corresponding calibration.
{\em XMM-Newton} pointed at the source on 2001 April 12--13 (rev.\ 246). The total exposure time was 36,714\,s.

The European Photon Imaging Camera (EPIC) onboard {\em XMM-Newton} carries three detectors: 
MOS1, MOS2 \citep{tur01} and pn \citep{str01}.
The MOS and pn observations were performed in Large Window and Full Frame Mode, respectively, all with a Medium filter.
We processed the data with the {\tt emproc} and {\tt epproc} tasks; high-background periods were 
removed by asking that the count rate of high-energy events ($E>10 \, \rm keV$) was less than 
0.35 and 0.40\,cts s$^{-1}$ on the MOS and pn detectors, respectively. The final exposure time was 
about 27--28\,ks for the MOS detectors, and about 12\,ks for the pn.
We extracted source counts from a circular region with 35 arcsec radius and background counts 
from a source-free circle with 70 arcsec radius.
We selected the best-calibrated single and double events only ({\tt PATTERN$<$=4}), and rejected events 
next to either the edges of the CCDs or bad pixels ({\tt FLAG==0}).
The absence of pile-up was verified with the {\tt epatplot} task.

   \begin{figure} 
  \centerline{\includegraphics[width=6.3cm,angle=-90]{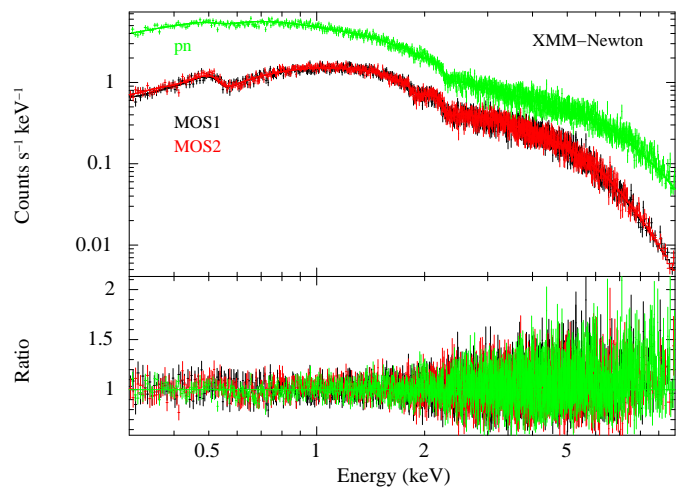}}
   \caption{X-ray spectra of \src{} acquired by the EPIC instrument onboard {\em XMM-Newton} on 2001 April 12--13. 
   Black, red and green symbols represent MOS1, MOS2 and pn data, respectively. The folded model, an absorbed power law 
   with $N_{\rm H} =3.30 \times 10^{20} \rm \, cm^{-2}$, is shown by solid lines of the same colour. The ratio between the data 
   and the folded model is plotted in the bottom panel.}
    \label{s0836:fig:xmmspec}
    \end{figure}

We grouped each spectrum with the corresponding background, redistribution matrix ({\tt RMF}), and 
ancillary ({\tt ARF}) files with the task {\tt grppha}, setting a binning of at least 25 counts for each spectral 
channel in order to use the chi-squared statistics.
The three spectra were analysed with {\sc Xspec} version 12.9.0, fitting them all together in the 
0.3--10\,keV energy range. 
We adopted a Galactic absorption column of $N_{\rm H} =2.76 \times 10^{20} \rm \, cm^{-2}$ from the 
LAB survey \citep{kal05} and the \citet{wil00} elemental abundances.

 \begin{table*}[!ht]     
 \begin{center}         
  \scriptsize
\caption{{\it XMM-Newton}-EPIC and {\it Swift}/XRT observation log and results of fitting the  spectra with power-law models.}         
\label{s0836:tab:xmmxrtspfits}    
 \begin{tabular}{ lrrrccccccc } 
 \hline 
 \hline 
 \noalign{\smallskip} 
  Sequence     & Start time  (UT)  & End time   (UT) & Exposure & $N_{\rm H}^a$  &  $\Gamma$  & $F^b$ &  $\chi^2_{\rm red}/$d.o.f.  &  $\Gamma^c$  & $F^{b,c}$&  $\chi^2_{\rm red}/$d.o.f.$^c$ \\ 
               & (yyyy-mm-dd hh:mm:ss)  & (yyyy-mm-dd hh:mm:ss)  &(s)       & & & &                        \\
 \noalign{\smallskip} 
 \hline 
 \noalign{\smallskip} 
%
%
\multicolumn{11}{c}{{\it XMM-Newton}/EPIC}\\
0112620101    &2001-04-12 17:36:00   &2001-04-13 03:47:54    &36714  & $3.3_{-0.2}^{+0.2}$ &  $1.340_{-0.007}^{+0.007}$ &  $4.69\pm0.03$      & $1.05/3509$  &         $1.325_{-0.004}^{+0.004}$ & $4.71\pm0.03$ &   $1.06/3510$ \\
 \noalign{\smallskip}
%
%
\multicolumn{11}{c}{{\it Swift}/XRT}\\
00036376046  &2015-10-30 01:16:27    &2015-10-30 03:03:07     &2936   &  $5.9_{-3.2}^{+3.7}$ &  $1.25_{-0.11}^{+0.12}$ &  $4.8\pm0.4$      & $1.233/58$  &         $1.17_{-0.06}^{+0.06}$ & $4.9\pm0.3$ &   $1.256/59$ \\
00036376048  &2015-11-01 02:51:36    &2015-11-01 23:59:54     &1923 &  $7.4_{-4.0}^{+4.8}$ &  $1.31_{-0.14}^{+0.15}$ &  $4.1\pm0.4$      & $0.909/34$  &                $1.18_{-0.08}^{+0.08}$ & $4.3\pm0.4$ &   $0.985/35$ \\
00036376051  &2015-11-03 06:12:03    &2015-11-03 23:50:55     &5062 &  $5.3_{-2.6}^{+3.0}$ &  $1.25_{-0.09}^{+0.09}$ &  $4.2\pm0.3$      & $0.966/89$  &         $1.18_{-0.05}^{+0.05}$ & $4.3\pm0.2$ &   $0.983/90$ \\
00036376052  &2015-11-05 01:17:50    &2015-11-05 12:28:53     &2329  &  $4.6_{-4.1}^{+5.0}$ &  $1.12_{-0.15}^{+0.16}$ &  $4.2\pm0.4$      & $0.824/35$  &         $1.07_{-0.09}^{+0.09}$ & $4.3\pm0.4$ &   $0.815/36$ \\
00036376053  &2015-11-07 15:20:12    &2015-11-07 18:34:53     &2941 &  $4.5_{-3.1}^{+3.6}$ &  $1.23_{-0.11}^{+0.12}$ &  $4.4_{-0.4}^{+0.3}$& $0.661/53$  &         $1.18_{-0.07}^{+0.07}$ & $4.5\pm0.3$ &   $0.665/54$ \\
total(046-053)&2015-10-30 01:16:27    &2015-11-07 18:34:53     &15192  & $6.3 _{-1.6}^{+1.7}$ &  $1.24_{-0.05}^{+0.05}$ & $4.4_{-0.1}^{+0.2}$ & $0.911/241$&         $1.14_{-0.03}^{+0.03}$ & $4.6_{-0.1}^{+0.2}$&   $0.970/242$  \\ 
 \noalign{\smallskip}
00036376054  &2015-11-10 16:37:03    &2015-11-10 18:12:38     &1976  & $3.1_{-3.1}^{+3.8}$  &  $1.10_{-0.13}^{+0.13}$ &  $5.4\pm0.5$      & $0.792/40$  &         $1.09_{-0.08}^{+0.08}$ & $5.4\pm0.4$ &   $0.774/41$  \\
 \noalign{\smallskip}
  \hline
  \end{tabular}
  \end{center}
  \tablefoot{$^a$:  $10^{20} \rm \, cm^{-2}$.
                  $^b$:  Observed flux in the 0.3--10 keV range ($10^{-11} \rm \, erg \, cm^{-2} \, s^{-1}$).  
                  $^c$: Fit performed with an absorption fixed to Galactic value ($N_{\rm H} =2.76 \times 10^{20} \rm \, cm^{-2}$). }
  \end{table*} 

The results of power-law fits with both free $N_{\rm H}$ and absorption fixed to the Galactic value, 
are reported in the first row of Table~\ref{s0836:tab:xmmxrtspfits}. The three spectra are shown in Fig.\ \ref{s0836:fig:xmmspec}.
In contrast with previous findings \citep{foschini06}, we did not find any evidence of substantial 
absorption beside the Galactic one. 
This discrepancy may be due to a combination of several factors. In particular, we adopted: 1) more recent calibration 
files and Science Analysis System ({\tt SAS}) software; 2) a different selection of the lower-energy bound for the spectral 
analysis; 3) a more conservative event selection to avoid periods when high-background flares could contaminate the data.
We also tried curved models (log-parabola and 
broken power law) to check for possible spectral curvature, but found none.

                 \subsection{Swift}  
%
%
The Neil Gehrels {\it Swift} Observatory \citep[][]{Gehrels2004} data (Target ID 36376) were collected by activating 
two dedicated target of opportunity observations (ToO) triggered as follow-up of AGILE  
detections \citep[][]{2015ATel.8223....1B, 2015ATel.8266....1P}.
The \sw\ data were processed and analysed by using standard procedures within 
the FTOOLS software (v6.17) and responses in the calibration database CALDB (20150731).

                 \subsubsection{Swift/XRT \label{s0836:data_xrt}}  

A log of all X-ray Telescope \citep[XRT, ][]{Burrows2005:XRT} 
observations triggered by our ToOs is reported in Table~\ref{s0836:tab:xmmxrtspfits}. 
The XRT data were collected for the most part in photon-counting (PC) mode 
and were processed with the {\sc xrtpipeline} (v.0.13.2). A moderate pile-up
affected the data, and it was corrected by adopting standard procedures 
\citep{vaughan2006:050315}, i.e.\ by determining the size of the 
core of the point spread function (PSF) affected  by pile-up by comparing 
the observed and nominal PSF, and excluding from the analysis all the events that fell 
within that region. The source events were thus extracted from an annulus with 
outer radius of 20 pixels (1 pixel $\sim2.36$\arcsec) and inner radius of 3 pixels. 
Background events were extracted from a circular source-free region nearby. 
Average spectra were extracted from each XRT observation, as well as 
the combined observations for the first ToO (046--053), 
and were binned to ensure at least 20 counts per energy bin,  
and fit in the 0.3--10\,keV energy range. 
   \begin{figure*}
  \vspace{-1truecm}
    \resizebox{\hsize}{!}{\includegraphics[angle=0]{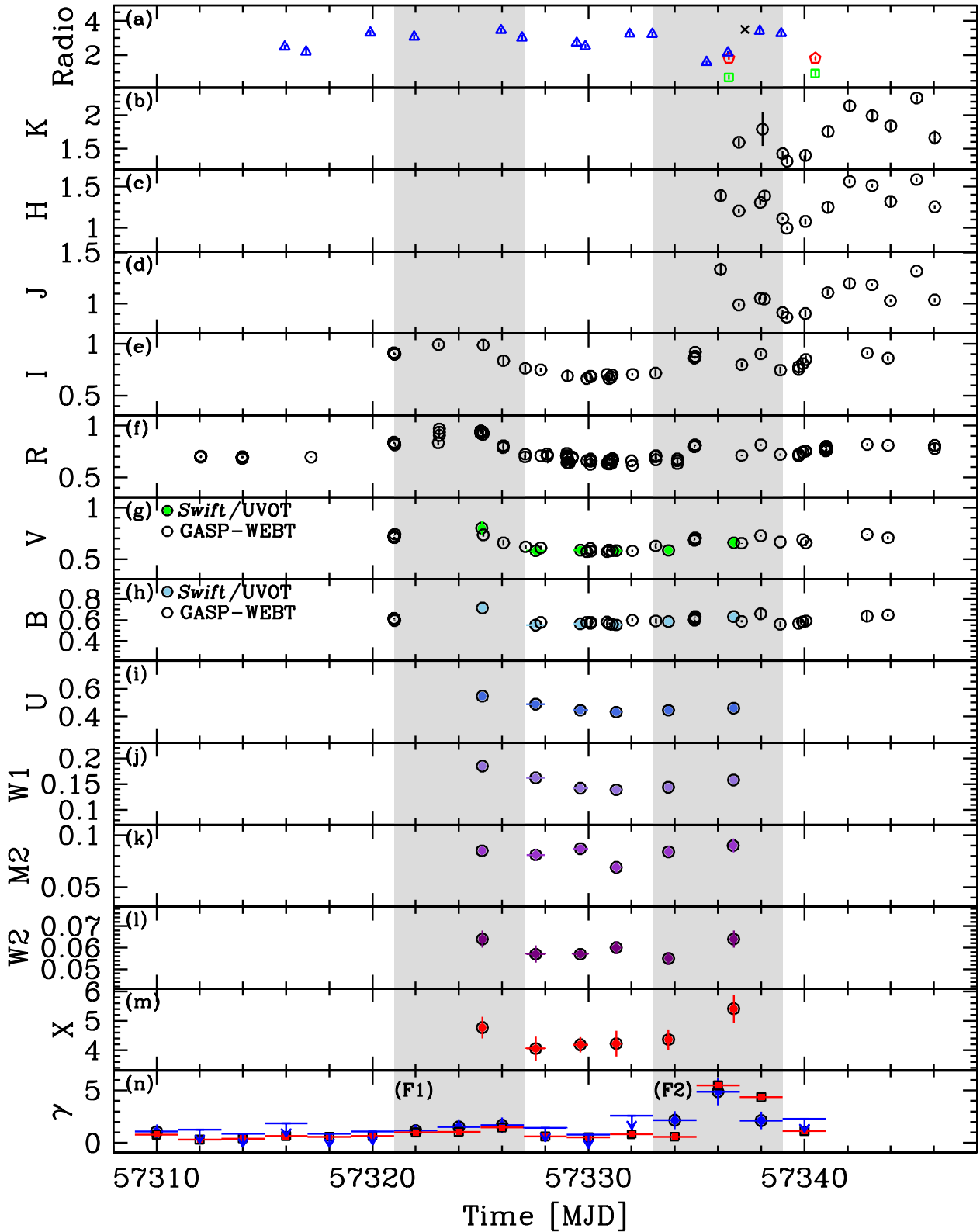}}
  \vspace{-1truecm}
    \caption{Multi-wavelength light-curves for the observing campaign on \src{}. 
    {\bf Panel (a)}: GASP-WEBT 5\,GHz (black cross sign), 37\,GHz (blue triangles), 86\,GHz (red diamonds),  and 228 \,GHz(green squares) data [Jy]. 
    {\bf Panels (b)--(h)}: $K$, $H$, $J$, $I$, $R$, $V$, $B$ bands (open circles, [mJy]).
    {\bf Panels (g)--(l)}: \sw{}/UVOT $v$, $b$, $u$, $w1$, $m2$, $w2$ bands (coloured discs, [mJy]). 
    {\bf Panel (m)}: \sw{}/XRT observed 0.3--10\,keV flux [$10^{-11}$\,\ferg{}]. 
    {\bf Panel (n)}: \agile{}/GRID (blue circles) and {\it Fermi}-LAT (red squares) data ($E>100$\,MeV, [$10^{-6}$\,\phcmsec]). 
   The grey-dashed areas mark the time-interval (F1, MJD 57321.0--57327.0; F2, MJD 57333.0--57339.0) used for accumulating the 
   almost simultaneous SEDs (orange (AGILE)/green (Fermi) and black(AGILE)/purple (Fermi) symbols, respectively) shown in Fig.~\ref{s0836:fig:sed}.}
    \label{s0836:fig:multi_lc}
    \end{figure*}
We adopted the same spectral models as those used for the  {\em XMM-Newton} data.
The results of power-law fits with both free $N_{\rm H}$ and absorption fixed to the Galactic value, 
are reported in Table~\ref{s0836:tab:xmmxrtspfits} and in Fig.~\ref{s0836:fig:multi_lc}, panel (m),
where we show the \sw{}/XRT (observed 0.3--10\,keV) light-curve derived with the free $N_{\rm H}$ model reported in Table \ref{s0836:tab:xmmxrtspfits}.

                 \subsubsection{Swift/UVOT \label{s0836:data_uvot}}  
 
The UV/Optical Telescope \citep[UVOT,][]{Roming2005:UVOT}  
observed \src\ simultaneously with the XRT in all optical and UV filters. 
The data analysis was performed using the {\sc uvotimsum} and 
{\sc uvotsource} tasks included in the FTOOLS. The latter 
task calculates the magnitude through aperture photometry within
a circular region and applies specific corrections due to the detector
characteristics. We adopted circular regions for source (5\,\arcsec radius) 
and background (10\,\arcsec radius). 
Fig.\ \ref{s0836:fig:multi_lc} shows the \sw{}/UVOT ($v$, $b$, $u$, $w1$, $m2$, $w2$ bands, panels (g)--(l)) light-curves.

                \subsection{GASP-WEBT \label{s0836:data_gasp}} 
                
%
Optical observations for the GASP-WEBT were done at the following observatories: 
Abastumani, Belogradchik, Calar Alto\footnote{Calar Alto data were acquired as part of the MAPCAT 
project: \href{http://www.iaa.es/~iagudo/\_iagudo/MAPCAT.html}{http://www.iaa.es/~iagudo/\_iagudo/MAPCAT.html}.}, 
Crimean, Mt.\ Maidanak, St.\ Petersburg, Roque de los Muchachos (LT), Rozhen, Southern Station of SAI, Teide (IAC80), 
and Astronomical Station Vidojevica. 
We calibrated the source magnitude with respect to the photometric sequences of \citet{vil97} and \citet{2014Ap.....57...30D}.
Light curves in the Johnson-Cousins' bands were carefully assembled.
Data scatter was reduced by binning data taken with the same telescope in the same night and by deleting clear 
outliers; in a few cases an offset with respect to the main trend was detected and corrected by shifting the whole dataset.
Near-infrared data in the $J$ band were acquired at the Campo Imperatore Observatory.
Millimetric radio data at 86 and 228 GHz were obtained at the IRAM 30\,m Telescope on 
Pico Veleta\footnote{IRAM 30\,m Telescope data were acquired as part of the POLAMI 
program: \href{http://polami.iaa.es}{http://polami.iaa.es}.} \citep[data calibrated as discussed in][]{2018MNRAS.474.1427A}.
Other radio data were taken at the Mets\"ahovi and Medicina\footnote{Operated by INAF--Istituto di Radioastronomia} (5 GHz) Observatories.

The $R$-band light-curve shows a well-defined maximum peaking at MJD=57322.5--57324.5 
(see Fig.~\ref{s0836:fig:multi_lc}), while another peak may 
have occurred at MJD=57335.5--57336.5, as confirmed by the $J$-band light curve.
 
  \section{Discussion \label{s0836:discussion}}

%
   \begin{figure*}
   \resizebox{\hsize}{!}
    {\includegraphics[angle=90]{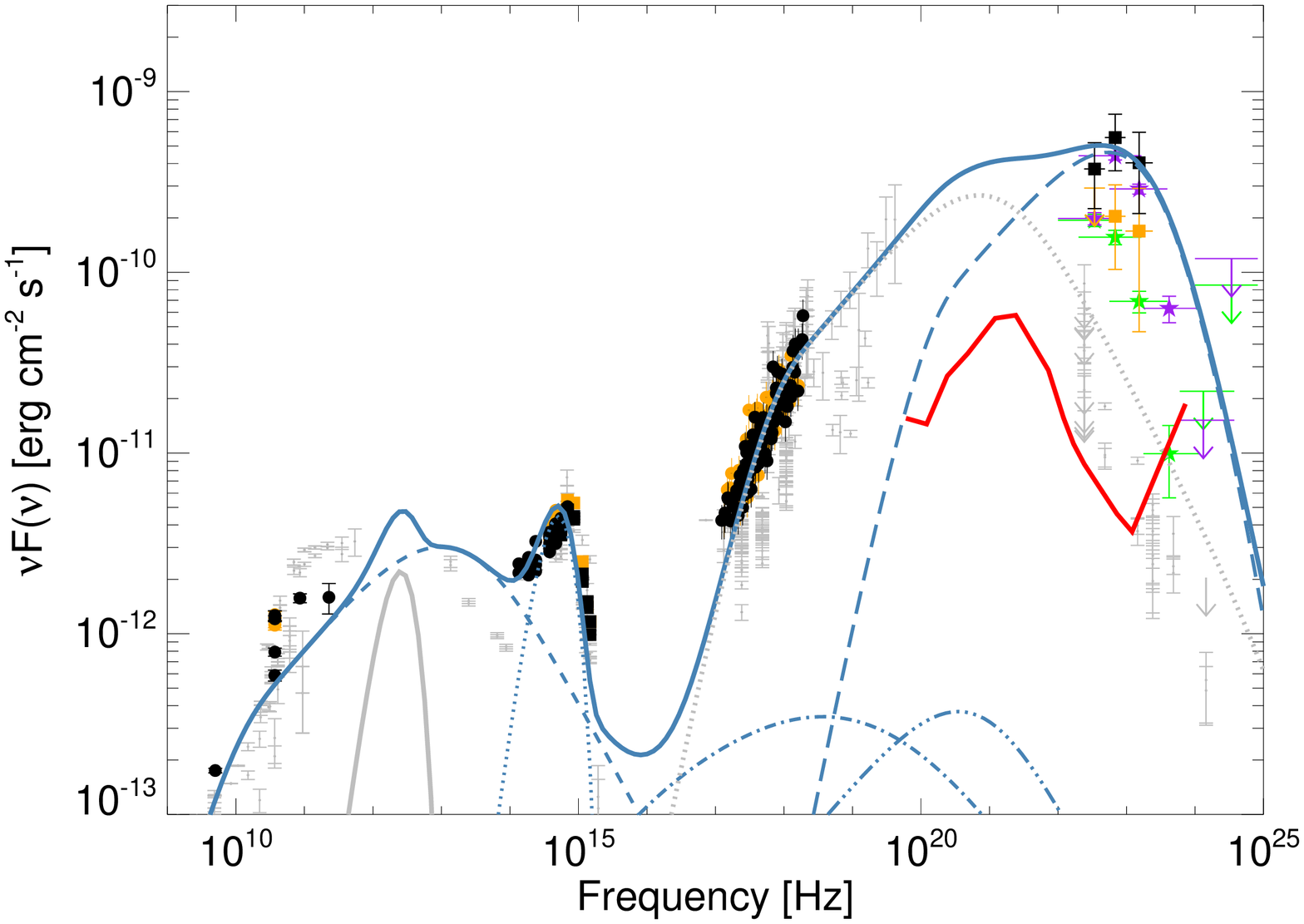}}
   \caption{Spectral energy distribution for the two flares (time intervals as reported in the caption of Fig.~\ref{s0836:fig:multi_lc}). 
   Orange symbols refer to the first flare, while black symbols to the second one AGILE data, while green and purple symbols
   refer to the {\it Fermi} first and second flare, respectively.
   Small grey points are archival data.  The blue lines represent the overall F2 SED fit (solid line) and each component, namely 
   the synchrotron emission (dashed line), the black-body approximation to the disc emission (dotted line), 
   the synchrotron self-Compton emission (SSC, dash-dotted line), 
   the external Compton emission off the disc (dash-triple-dot line), and the external Compton emission off the 
   broad-line region (long-dashed line), respectively. The light-grey solid and dotted lines represent the torus and the 
   external Compton emission off the torus photons, respectively. The red curve represents the e-ASTROGAM sensitivity 
   for an integration time of 6 days (comparable to the AGILE {and {\it Fermi}} integration time for the spectral analysis).}
    \label{s0836:fig:sed}
    \end{figure*}

Our observing campaign allowed us to collect multi-wavelength (MWL) data covering the two main \gray{} flares detected by AGILE.
In particular, the second and most prominent \gray{} flare (F2) has a much richer MWL coverage, as shown in 
Fig.~\ref{s0836:fig:multi_lc}, including the mm and the near infra-red wavelength.

In order to build the spectral energy distribution, we corrected the near-infrared, optical, and UV data for Galactic 
reddening by adopting a value of 0.083 mag in the Johnson's $V$ band (from NED) and the \citet{1989ApJ...345..245C} 
mean extinction laws, with the parameter $R_V=3.1$, the standard value for the diffuse interstellar medium.
Since these laws present a bump at 2175 \AA, they were convolved with the filters' effective areas to derive 
the reddening correction for the \sw{}/UVOT bands~\citep[see][for further details]{2010A&A...524A..43R}. 

Fig.~\ref{s0836:fig:sed} shows the spectral energy distribution for the two flares of \src{} (time intervals as reported in the caption 
of Fig.~\ref{s0836:fig:multi_lc}). 
Orange and black symbols refer to the AGILE first and second flares, respectively, 
while green and purple symbols refer to the {\it Fermi} first and second flare, respectively.
Small grey points are archival data provided by the ASI/ASDC {\it SED Builder Tool}~\citep{2011arXiv1103.0749S}.

The data show not only the typical double-humped shape of the blazar SED, but also a prominent disc bump
peaking in the UV energy band. Moreover, while the rising branch of the inverse Compton, the disc and, only marginally, the poorly
constrained synchrotron emission are almost consistent with the non-simultaneous data (grey points), the high-energy peak ($E>100$\,MeV) 
is about one order of magnitude more intense with respect to the archival ones.

In order to model the SED, we took into account a one-zone leptonic model.
The emission along the jet is assumed to be produced in a spherical blob with co-moving radius $R_{\rm blob}$ by
accelerated electrons characterised by a broken power-law particle density distribution,
\begin{equation}
n_{e}(\gamma)=\frac{K\gamma_{\rm b}^{-1}} 
{(\gamma/\gamma_{\rm b})^{\alpha_{\rm l}}+
(\gamma /\gamma_{\rm b})^{{\alpha}_{\rm h}}}\,,
\label{eq:ne_gamma}
\end{equation}
where $\gamma$ is the electron Lorentz factor varying between $20<\gamma<5\times10^{3}$, $\alpha_{\rm l}$ and
$\alpha_{\rm h}$ are the pre-- and post--break electron distribution spectral indices, respectively, and $\gamma_{\rm b}$ is the
break energy Lorentz factor. We assume that the blob contains an homogeneous magnetic field
$B$ and that it moves with a bulk Lorentz factor
$\Gamma$ at an angle $\Theta_{0}$  with respect to the line of sight. The relativistic Doppler factor is 
$\delta = [ \Gamma \,(1 - \beta \, \cos{\Theta_{0}})]^{-1}$, where $\beta$ is the blob bulk speed in units of the speed of light.
Our assumed and best fit parameters are listed in Table~\ref{s0836:tab:sedfit}.

Our modelling of the \src{} high-energy emission is based on an inverse Compton  (IC) model with three main sources of 
external seed photons.

The first one is the accretion disc characterised by a blackbody spectrum peaking in the UV with a bolometric luminosity  $L_{\rm d}$
for an IC-scattering blob at a distance $z_{\rm jet}$  from the central part of the disc. 
{The value of the black hole mass is based on the one computed by \citet{2015ApJ...807..167T}, 
$M_{\rm BH} = 5\times 10^{9}$\,M$_{\odot}$. From \citet{2014MNRAS.442..629R} we can approximate 
the disc luminosity to $L_{\rm d} \approx 2.2\times10^{47}$\,\ergsec
and $T_{\rm d} \approx 3 \times 10^{4}$ $^{\circ}K$, computed by means of a black-body approximation of the disc model.
Moreover, we assume the bulk Lorentz factor $\Gamma = 20$ similar to the value reported in~\cite{2015ApJ...804...74P}. 
Our almost simultaneous data marginally sample the synchrotron component of the SED, preventing us to 
to derive additional constraints from this portion of the SED.

Almost simultaneous SEDs were previously modelled by \citet{2010MNRAS.405..387G} and \citet{2015ApJ...807..167T} in different emission states.
We base our assumed parameters on those derived by these authors. In particular, we fix $z_{\rm jet} = 1 \times 10^{18}$\,cm 
(distance from the central black-hole at which the emission takes place, intermediate between the values reported in the cited references); this is 
to be compared with $r_{\rm s} = 2GM/c^{2} \approx 1.5\times10^{15}$\, cm. 

The second source of external seed photons is the broad-line region, placed at a distance from the central black-hole of
$R_{\rm BLR} \approx 1.6\times 10^{18}$\,cm, and assumed to reprocess $f_{\rm BLR} = 3\%$ of the irradiating continuum 
(obtained by considering a BLR cloud-coverage factor of 30\%  and a 10\% reflectivity factor for each single cloud).
Given the relative locations of the disc, the emitting blob, and the BLR, we consider disc photons entering the blob 
from behind (therefore de-boosted) while the BLR photons may be considered isotropic within $R_{\rm BLR}$ in the source 
frame~\citep[head-on, hence boosted, see][]{1994ApJS...90..945D}.

The third source of external seed photons is the dusty torus, assumed to be located at a distance of $R_{\rm Torus} \approx 10^{19}$\,cm,
emitting in the infra-red energy range with $T_{\rm Torus} \approx 100$ $^{\circ}K$ and assumed to reprocess a fraction $f_{\rm Torus} = 50\%$ of the irradiating continuum.

 \begin{table} 	
 \begin{center} 	
\caption{Parameters for the second flare (F2) SED model. $\Gamma$, $L_{\rm d}$, and $T_{\rm d}$ are assumed as fixed
parameters.}
 \label{s0836:tab:sedfit} 	
 \begin{tabular}{ lll } 
 \hline 
 \hline 
 \noalign{\smallskip} 
  Parameter    & Value  & Unit   \\ 
 \noalign{\smallskip} 
   \hline 
 \noalign{\smallskip} 
\multicolumn{3}{c}{{\it Fixed parameters}}\\
$z_{\rm jet}$ &       	         $1 \times 10^{18}$			& cm    \\
$R_{\rm blob}$ &       	$4 \times 10^{16}$			& cm    \\
$L_{\rm d}$ &   			$2.2 \times 10^{47}$		    	& \ergsec   \\
$T_{\rm d}$ &    		$3 \times 10^{4}$   			& $^{\circ}K$   \\
$R_{\rm BLR}$ &       	$1.6\times 10^{18}$           	& cm    \\
$f_{\rm BLR}$ &       	         $3$			                         & \%    \\
$R_{\rm Torus}$ &       	$1\times 10^{19}$           	& cm    \\
$T_{\rm Torus}$ &    		$1 \times 10^{2}$   			& $^{\circ}K$   \\
$f_{\rm Torus}$ &       	         $50$			                         & \%    \\
$\Theta_{0}$  &       		$2$						&  degrees  \\
$\Gamma$ &       		$20$						&    \\
$\delta$ &       			$27$					        &    \\

 & &  \\
 \multicolumn{3}{c}{{\it Best fit parameters}}\\
$\alpha_{\rm l}$  &       	$2.1$					&    \\
$\alpha_{\rm h}$ &       	$5$					&    \\
$\gamma_{\rm min}$ &     $20$						&    \\
$\gamma_{\rm b}$ &         $750$					&    \\
$K$ &       				$19$		          			& cm$^{-3}$   \\
$B$ &       				$0.9$					& G    \\
 \noalign{\smallskip}
  \hline
  \end{tabular}
  \end{center}
  \end{table} 

Fig.~\ref{s0836:fig:sed} shows the overall SED fit (solid line) and each component, namely the synchrotron emission (dashed line), 
the thermal disc emission (dotted line), the synchrotron self-Compton emission (SSC, dash-dotted line), the external Compton emission off the disc 
(dash-triple-dot line), the external Compton emission off the broad-line region (long-dashed line), the infra-red torus (light-grey solid line), and the
external Compton emission off the torus photons (light-grey dotted lines), respectively.
Fig.~\ref{s0836:fig:sed} clearly shows how the Compton part of the spectrum dominates over the synchrotron and the thermal ones.

The energetics of \src{} can be computed by estimating the isotropic luminosity in the \gray{} energy band, $L_{\gamma}^{\rm iso}$.
For a given source with redshift $z$, the isotropic emitted luminosity in an energy band  can be computed following~\citet{2010ApJ...712..405V}.
Using the observed \gray{} flux and photon index for the second flare labelled F2 (see Table~\ref{s0836:tab:fgamma}), 
for $E>100$\,MeV we obtain $L_{\gamma, E>100{\rm MeV}}^{\rm iso} \approx 3\times10^{49}$\,\ergsec{}, 
while the Eddington luminosity is $L_{\rm Edd} \approx 6\times 10^{47}$\,\ergsec{}
(implying $L_{\rm d}/L_{\rm Edd} \sim 0.3$).
We can define the total power carried in the jet, $P_{\rm jet}$ following \citet{2001MNRAS.327..739G} as
\begin{equation}
P_{\rm jet} = P_{\rm B} + P_{\rm p} + P_{\rm e} + P_{\rm rad},
\label{eq:Pjet}
\end{equation}
where $P_{\rm B}$, $P_{\rm p}$, $P_{\rm e}$, and $P_{\rm rad}^{\rm bol}$ 
are the power carried by the magnetic field, the cold protons,
the relativistic electrons, and the produced radiation, respectively.
In order to compute the different components, we use the formalism presented in \citet{2008MNRAS.385..283C}.
Including the counter-jet contribution, we obtain 
$P_{\rm B} \approx 3\times10^{45}$\,erg\,s$^{-1}$, 
$P_{\rm e} \approx 6.2\times10^{45}$\,erg\,s$^{-1}$,
$P_{\rm p} \approx 2.1\times10^{47}$\,erg\,s$^{-1}$, 
$P_{\rm rad}^{\rm bol} \approx 1.8\times10^{47}$\,erg\,s$^{-1}$, 
which yields $P_{\rm jet} \approx 4\times10^{47}$\,erg\,s$^{-1}$.

Alternatively, we may estimate the jet power following \cite{2015MNRAS.451..927Z}, 
$P_{\rm jet} \approx 1.3 (\eta/0.2) \dot{M}c^{2} \approx 1.43\times10^{48}$\,erg\,s$^{-1}$, 
assuming $\dot{M}\approx L_{\rm d}/(\eta c^{2})$, $\eta=0.3$, and $L_{\rm d} = 2.2\times10^{47}$\,erg\,s$^{-1}$.
Combining this results with the previous one, we obtain that the jet power is in the range (0.4--1.4)$\times10^{48}$\,erg\,s$^{-1}$.

The study of high-redshift blazars has a strong impact on cosmology providing crucial information about 
i) the formation and growth of super-massive BHs, ii) the connection between the jet and the central engine, 
and iii) the role of the jet in the feedback occurring in the host galaxies \citep{2011MNRAS.416..216V}.
Recent hard X-ray surveys \citep{2009ApJ...699..603A, 2010MNRAS.405..387G, 2012ApJ...751..108A} demonstrated 
to be more effective in detecting high-redshift blazars compared to \gray{} surveys. The main reason is that the 
SEDs of these sources peak in the MeV region and detection becomes a difficult task for \gray{} instruments.

In this context, the e-ASTROGAM mission~\citep[proposed as a medium size mission in the ESA M5 call,][]{2017ExA....44...25D} will 
have a great potential in detecting these blazars taking advantage of its soft \gray{} band (0.3--3000)\,MeV as compared 
with AGILE and \fermi{}, with a sensitivity in the energy range 1--10\,MeV more than one order of magnitude better with respect to COMPTEL.
In Fig.~\ref{s0836:fig:sed} is reported, as a red line, the e-ASTROGAM extra-galactic sensitivity for an integration time of 6 days, comparable
to the AGILE integration time during the \gray{} flares. We can appreciate the e-ASTROGAM excellent performance in detecting such kind of objects, 
providing crucial information both in the rising of the inverse Compton energy range and at its peak.
High-redshift sources should have high accretion rates, close to the Eddington limit, yielding to high Compton dominance 
($\approx$100 in \src{}). 
e-ASTROGAM will substantially advance our knowledge of MeV blazars up to redshift $z = 4.5$, with implications for blazar physics, cosmology,
and the study of both the extra-galactic background light and the inter-galactic magnetic field.
These observations will be invaluable and complementary to ATHENA~\citep{ASNA:ASNA201713323} data for the study of 
super-massive black holes with the possibility to investigate how the two populations of AGNs (radio-quiet and radio-loud) evolve with redshift.

  \section{Conclusions \label{s0836:conclusions}}
In this paper we presented the almost simultaneous data collected on the high-redshift flat-spectrum radio quasar \src{}. 
The AGILE, {\it Fermi}-LAT, \sw{}, and GASP-WEBT data allowed us to investigate both the non-thermal emission
originating from the jet emerging from the central black-hole and the properties of the \gray{} emitting region.
We found that the data collected in our observing campaign can be modelled
by means of a simple one-zone leptonic model with the emission zone within the broad-line region. 
Moreover, we discussed how such class of sources might be suitable candidates for the proposed e-ASTROGAM \gray{} mission.

  \begin{acknowledgements} 
  
      We thank the referee for their very constructive comments, which helped in substantially improve the manuscript.
      {\it AGILE} is an ASI space mission developed with programmatic support by INAF and INFN. We acknowledge partial support through the ASI 
      grant no. I/028/12/0.
      SV and PR acknowledge contract ASI-INAF I/004/11/0 and INAF/IASF Palermo were most of the work was carried out. 
      SV acknowledges financial contribution from the agreement ASI-INAF n.2017-14-H.0.
      Part of this work is based on archival data, software or online 
      services provided by the ASI SPACE SCIENCE DATA CENTER (ASI-SSDC).
      SV and PR thank Leonardo Barzaghi and Sara Baitieri for useful discussions.
      The Osservatorio di Torino team acknowledges financial contribution from the agreement 
      ASI-INAF n. 2017-14-H.0 and from the contract PRIN-SKA-CTA-INAF 2016.
      OMK acknowledges financial support by the by Shota Rustaveli National Science Foundation under contract FR/217950/16
      and grants NSFC11733001, NSFCU1531245.
      IA acknowledges support by a Ram\'on y Cajal grant of the Ministerio de Econom\'ia y Competitividad (MINECO) of Spain. 
      The research at the IAA--CSIC was supported in part by the MINECO through grants AYA2016--80889--P, AYA2013--40825--P, 
      and AYA2010--14844, and by the regional government of Andaluc\'{i}a through grant P09--FQM--4784. 
      IRAM is supported by INSU/CNRS (France), MPG (Germany) and IGN (Spain). Calar Alto Observatory 
      is jointly operated by the MPIA and the IAA-CSIC.
      This research was partially supported by the Bulgarian National Science Fund of the Ministry of Education and Science under grant DN~08-1/2016.
      The St.Petersburg University team acknowledges support from Russian Science Foundation grant 17-12-01029.
       AZT-24 observations are made within an agreement between  Pulkovo, Rome and Teramo observatories.
      GD and OV gratefully acknowledge the observing grant support from the Institute of Astronomy and Rozhen National Astronomical Observatory,
      Bulgaria Academy of Sciences, via bilateral joint research project "Observations of ICRF radio-sources visible in optical domain" 
      (the head is G. Damljanovic). This work is a part of the Projects No 176011 ("Dynamics and kinematics of celestial bodies and systems"), 
      No 176004 ("Stellar physics") and No 176021 ("Visible and invisible matter in nearby galaxies: theory and observations") supported by the 
      Ministry of Education, Science and Technological Development of the Republic of Serbia.
      The Maidanak Observatory team acknowledges support from Uzbekistan Academy of Sciences grants N. F2-FA-F027 and F.4-16.
  \end{acknowledgements}
\clearpage

\bibliographystyle{aa} 

%


\end{document}